\newcounter{myctr}
\def\myitem{\refstepcounter{myctr}\bibfont\noindent\ifnum\themyctr>9\else\phantom{0}\fi\hangindent17pt\themyctr.\enskip}

\documentclass{ws-ijqi}

\begin{document}

\markboth{T. Prud\^{e}ncio}
{No-cloning theorem in thermofield dynamics}

\catchline{}{}{}{}{}

\title{NO-CLONING THEOREM IN THERMOFIELD DYNAMICS}

\author{T. PRUD\^ENCIO}

\address{Instituto de F\'\i sica, Universidade de Bras\'ilia - UnB, CP: 04455, 70919-970, Bras\'ilia - DF, Brazil.\\
International Institute of Physics, Universidade Federal do
Rio Grande do Norte, Av. Odilon Gomes de Lima, 1722, 59078--400, Natal, RN, Brazil.}

\maketitle


\begin{abstract}
We discuss the relation between the no-cloning theorem from quantum information and the doubling procedure used in the formalism of thermofield dynamics (TFD). We also discuss how to apply the no-cloning theorem in the context of thermofield states defined in TFD. Consequences associated to mixed states, von Neumann entropy and thermofield vacuum are also addressed.
\end{abstract}


\section{Introduction}	

In 1975, Takahashi and Umezawa\cite{takahashi} proposed a formalism based on operators called thermofield dynamics (TFD) in order to describe quantum field theory at finite temperature by means of a structure of Hilbert space. This approach was not totally new since a similar idea appeared in a paper by Araki and Woods, in 1963, considering the case of a free boson gas\cite{araki}. Takahashi and Umezawa's approach is an alternative to the density operator formalism in quantum statistical mechanics proposed by Landau\cite{landau} and von Neumann\cite{neumann}$^{,}$\cite{neumannb} in 1925 and to the finite temperature quantum field theory proposed by Matsubara in 1955\cite{matsubara}. In fact, TFD is an operator-algebraic approach to quantum statistical mechanics\cite{landsman} and a real time formalism to finite temperature quantum field theory\cite{matsumoto2}$^{,}$\cite{das}.

The basic ingredients of TFD are the doubling of freedom degrees in the Hilbert space $\mathcal{H}$ which describes the physical system and the building of a finite temperature vacuum, called hereafter thermofield vacuum, by means of a Bogoliubov transformation realized in the zero temperature vacuum state defined into the Hilbert space $\mathcal{H}\otimes\tilde{\mathcal{H}}$, also called Liouville space \cite{tay}$^{,}$\cite{ban}, where the space $\tilde{\mathcal{H}}$ is constructed by means of a tilde conjugation rule $\mathcal{H}\rightarrow \tilde{\mathcal{H}}$\cite{santanab}. This procedure is made in such a way that the expectation value of any operator from $\mathcal{H}$ in the thermofield vacuum coincides with the statistical mean value.

TFD approach has been largely applied in the study of finite temperature systems, from high energy physics \cite{umezawab2}$^{,}$\cite{kobes}$^{,}$\cite{balachandran}$^{,}$\cite{costa}$^{,}$\cite{leineker} to quantum optics \cite{knight}$^{,}$ \cite{barnett}$^{,}$\cite{mann}$^{,}$\cite{chaturvedi}$^{,}$\cite{vourdas}$^{,}$\cite{matrasulov} and condensed matter physics \cite{umezawab}$^{,}$\cite{suzuki}$^{,}$\cite{matsumoto4}, in different contexts as gauge theories \cite{ojima}, Anderson model\cite{matsumoto4}, renormalization group\cite{nakano1}, Casimir effect\cite{silva}$^{,}$\cite{belich}, string field theory\cite{leblanc}, supersimmetry\cite{nakahara}, noncommutative theories\cite{balachandran}$^{,}$\cite{costa}, non-classical states\cite{chaturvedi}, thermal and quantum fluctuations\cite{yamanaka}, master equations\cite{tay}, among others \cite{santanab}.

However, quantum information theory was not still explored in a TFD context, although its potential to deal with quantum information protocols in systems at finite temperature \cite{huyet}$^{,}$\cite{santana}$^{,}$\cite{brune}$^{,}$\cite{almeida}.

In quantum information the storage properties of qubits from quantum computers in comparison to bits from classical computers can be viewed as a result of a Hilbert space structure in quantum mechanics, that leads in the quantum computer case to strong processing results\cite{ladd}$^{,}$\cite{gotesman}. In this sense, we can also explore the Hilbert state structure from TFD approach in order to trace some parallels with achievements from quantum information.

The no-cloning theorem was proved, in 1982, by Wooters and Zurek\cite{wooters} and Dieks\cite{dieks} to pure states, and then to mixed states by Barnum \textit{et al.}\cite{barnum}, in 1996, and to entangled states by Koashi and Imoto\cite{koashi}, in 1998, leading to some other consequences as protocols of quantum cloning\cite{scarani}$^{,}$\cite{buzek}$^{,}$\cite{linares} and telecloning\cite{murao}$^{,}$\cite{koike}.

In this paper we discuss the relation between the no-cloning theorem and the doubling procedure used in the formalism of TFD. We also discuss how to apply the no-cloning theorem in the context of thermofield states defined in TFD. Consequences associated to states in $\mathcal{H}\otimes\tilde{\mathcal{H}}$, mixed states, von Neumann entropy and thermofield vacuum are also addressed. 

In order to clarify notation and subject, in the next section we review some basic aspects of TFD approach.

\section{Thermofield approach}

Consider an operator $\hat{A}$ acting on a Hilbert space $\mathcal{H}$ generated by Fock states $|n\rangle$. Its expectation value in a given ensemble is expressed by
\begin{eqnarray}
\langle \hat{A}\rangle = Tr(\hat{\rho} \hat{A}), 
\end{eqnarray}
where $\hat{\rho}$ is the density operator in the corresponding ensemble. In thermofield dynamics this expectation value is evaluated by means of the definition of a thermofield vacuum $
|0(\beta )\rangle $, where $\beta=1/T$ is the inverse of temperature $T$ ($k_{B}=\hbar=1$), giving the same result as statistical approach, i.e.,
\begin{eqnarray}
\langle 0(\beta )|\hat{A}|0(\beta )\rangle = Tr(\hat{\rho} \hat{A}). 
\label{med}
\end{eqnarray}
As a consequence, the thermal vaccuum state $|0(\beta )\rangle$ is associated to the density
operator $\hat{\rho}$. For this reason we need to describe it in a Hilbert space larger than the Hilbert space $\mathcal{H}$ generated by the Fock states $|n\rangle$. Then, the thermofield vacuum $|0(\beta )\rangle$ is not a vector state in the Hilbert space $\mathcal{H}$ described by the Fock states $|n\rangle$, but a state in another enlarged Hilbert space $\mathcal{H}\otimes\tilde{\mathcal{H}}$, where $\tilde{\mathcal{H}}$ is the Hilbert space conjugated to $\mathcal{H}$. In fact, in order to describe $|0(\beta)\rangle $ as a vector state, we need to double the degrees of freedom of the Hilbert space $\mathcal{H}$ by a formal procedure named tilde conjugation \cite{santanab}, creating the space $\mathcal{H}\otimes\tilde{\mathcal{H}}$.

In order to construct the thermofield vacuum in the space $\mathcal{H}\otimes\tilde{\mathcal{H}}$, let us suppose we can describe $|0(\beta)\rangle$ in terms of the Fock state basis $|n\rangle$ and unknown vectors $|c_{n} \rangle \in \tilde{\mathcal{H}}$ by means of the following expansion
\begin{eqnarray}
|0(\beta)\rangle =\sum_{n}|c_{n} \rangle|n\rangle.
\label{evt}
\end{eqnarray}
The next step is to find the form of the unkown vectors $|c_{n}\rangle$ such that the relation (\ref{evt}) turns out to be true.

Consider the system described by a thermal equilibrium density matrix
\begin{eqnarray}
\hat{\rho}=e^{-\beta \hat{H}}/Z,
\end{eqnarray}
where $Z=Tr(e^{-\beta \hat{H}})$ is the partition function and the energy spectrum $E_{n}$ of the hamiltonian $\hat{H}$,
\begin{eqnarray}
\hat{H}|n\rangle = E_{n}|n\rangle.
\end{eqnarray}
Then the expectation value of $\hat{A}$ in (\ref{med}) can be written as
\begin{eqnarray}
Tr(\hat{\rho} \hat{A})=\frac{1}{Z}\sum_{n}e^{-\beta E_{n}}\langle n|\hat{A}|n\rangle.
\label{rodado}
\end{eqnarray}
We can also write the same in terms of $|0(\beta )\rangle$ by using equations (\ref{med}), (\ref{evt}) and the assumption that $\hat{A}$ does not act on $|c_{n} \rangle$ vectors,
\begin{eqnarray}
\langle 0(\beta )|\hat{A}|0(\beta )\rangle =\sum_{m,n}\langle c_{m}|c_{n} \rangle\langle m|\hat{A}|n\rangle.
\label{somado}
\end{eqnarray}
Comparing equations (\ref{rodado}) and (\ref{somado}), we find
\begin{eqnarray}
\langle c_{m}|c_{n} \rangle = \frac{1}{Z}e^{-\frac{1}{2}\beta(E_{m} + E_{n})}\delta_{mn}.
\end{eqnarray}
Thus, by defining the Fock states $|\tilde{n}\rangle \in \tilde{\mathcal{H}}$ a basis product can be given to $\mathcal{H}\otimes\tilde{\mathcal{H}}$ and we can write $|c_{n} \rangle$ vectors as
\begin{equation}
|c_{n} \rangle = \frac{1}{\sqrt{Z}}e^{-\frac{1}{2}\beta E_{n}}|\tilde{n}\rangle.
\label{vec1}
\end{equation}
The vector in equation (\ref{vec1}) satisfy the conditions (\ref{med}) and (\ref{evt}) and we can finally express the thermofield vacuum in terms of $|n,\tilde{n}\rangle \in \mathcal{H}\otimes\tilde{\mathcal{H}}$,
\begin{eqnarray}
|0(\beta )\rangle =\frac{1}{\sqrt{Z}}\sum_{n}e^{-\beta E_{n}/2}|n,\tilde{n}\rangle.
\end{eqnarray}
We can also have operators acting on $|\tilde{n}\rangle \in \tilde{\mathcal{H}} $. We distinguish them by putting a tilde on top of a capital letter, e.g., $\tilde{A}$. In TFD, tilde conjugation rules \cite{matsumoto3} realize a mapping between $\hat{A}$ operators acting on $|n\rangle$ and $\tilde{A}$ acting on $|\tilde{n}\rangle$. These rules are summarized by
\begin{eqnarray}
\widetilde{(\hat{A}\hat{B})}&=&\tilde{A}\tilde{B},\\
\widetilde{(z\hat{A}+w\hat{B})}&=& z^{\ast }\tilde{A}+w^{\ast }\tilde{B},\\
\widetilde{(\hat{A}^{\dagger })}&=&(\tilde{A})^{\dagger },\\
\widetilde{(\tilde{A})}&=& \pm\hat{A}, \label{pm}\\
\lbrack \hat{A},\tilde{B}]_{\pm}&=&0, \label{pm2}
\end{eqnarray}
where the operators $\hat{A}$ and $\hat{B}$ act only in the Hilbert space spanned
by $|n\rangle$, and $\tilde{A}$ and $\tilde{B}$ act only in the Hilbert
space generated by $|\tilde{n}\rangle $, where $z$ and $w$ are complex
numbers, $z^{\ast }$ and $w^{\ast }$ are their respective complex conjugated. In equation (\ref{pm}), $+$ is for bosons and $-$ is for fermions \cite{ojima}. In equation (\ref{pm2}) $+$ means commutation for bosons and $-$ is anticommutation for fermions \cite{matsumoto3}.

From the operators $\hat{A}$ and $\tilde{A}$ it is possible to derive also other operators that will act on the total space $\mathcal{H}\otimes\tilde{\mathcal{H}}$. We can use a bar to distingish such operators, following notation similar to Ojima's\cite{ojima}, and write generally
\begin{eqnarray}
\bar{A}= \bar{A}(\hat{B}_{1}, ..., \hat{B}_{n}, \tilde{C}_{1}, ...,\tilde{C}_{m})
\label{bar1}
\end{eqnarray}
where here we have $\bar{A}$ as a function of $\hat{B}_{1}, ..., \hat{B}_{n}$ and $\tilde{C}_{1},..., \tilde{C}_{m} $. For instance, we could have
\begin{eqnarray}
\bar{A} = z\hat{A} + w\tilde{A}.
\label{bar}
\end{eqnarray}
where $z$ and $w$ are complex numbers. Such structure of operators is in fact rigorous as a mathematical formulation of a thermal theory\cite{santkhan95} and consequently we have a important relationship between thermofield vacuum and density operators to each given temperature $T=\beta^{-1}$,
\begin{eqnarray}
|0(\beta )\rangle \longrightarrow \hat{\rho} =\frac{e^{-\beta \hat{H}}}{Z}.
\end{eqnarray}
In the Liouville space $\mathcal{H}\otimes\tilde{\mathcal{H}}$ given by TFD, the zero temperature vacuum state is given by $|0,\tilde{0}\rangle $, whose corresponding density operator at zero temperature is $\hat{\rho}_{0} =|0\rangle \langle 0|$.
By applying a Bogoliubov transformation $e^{-i\bar{G}}$ on this vacuum state $|0,\tilde{0}\rangle$, the thermofield vacuum is generated at a finite temperature $T=\beta^{-1}$,
\begin{eqnarray}
|0(\beta )\rangle =e^{-i\bar{G}}|0,\tilde{0}\rangle,
\label{adk}
\end{eqnarray}
where $\bar{G}$ is an unitary operator mixing $|n\rangle \in \mathcal{H}$ and $|\tilde{n}\rangle \in \tilde{\mathcal{H}}$ 
by acting on the Hilbert space $\mathcal{H}\otimes\tilde{\mathcal{H}}$, similarlly to (\ref{bar1}) and (\ref{bar}), 
but with a two-mode squeezing operator form \cite{santana},
\begin{eqnarray}
\bar{G}=i\theta(\beta)\left(\hat{a}^{\dagger}\tilde{b}^{\dagger} - \hat{a}\tilde{b} \right)
\label{bog}
\end{eqnarray}
where $\theta=\theta(\beta)$ is a parameter related to a thermal distribution, $\hat{a}^{\dagger}$ and $\tilde{b}^{\dagger}$ are creation operators acting on spaces $\mathcal{H}$ and $\tilde{\mathcal{H}}$, and 
$\hat{a}$ and $\tilde{b}$ are annihilation operators acting on spaces $\mathcal{H}$ and $\tilde{\mathcal{H}}$, respectively. 
In the case of a bosonic oscillator system, we have $\theta=\tanh^{-1}(e^{-\beta\omega/2})$, related to a Bose-Einsten statistics, 
and for a fermionic oscillator system we have $\theta=\tan^{-1}(e^{-\beta\omega/2})$, 
related to a Fermi-Dirac statistics. 

From Bogolioubov transformation (\ref{bog}), we can define at given finite temperature $T=\beta^{-1}$, thermofield operators from the creation and annihilation operators
\begin{eqnarray}
\bar{a}_{\beta}^{\dagger } &=& e^{-i\bar{G}}\hat{a}^{\dagger }e^{i\bar{G}}, \label{lyiul1}\\
\bar{a}_{\beta} &=& e^{-i\bar{G}}\hat{a}e^{i\bar{G}} \label{lyiul2},
\end{eqnarray}
\begin{eqnarray}
\bar{b}_{\beta}^{\dagger } &=& e^{-i\bar{G}}\tilde{b}^{\dagger }e^{i\bar{G}}, \label{lyiul3}\\
\bar{b}_{\beta} &=& e^{-i\bar{G}}\tilde{b}e^{i\bar{G}}. \label{lyiul4}
\end{eqnarray}
These operators commute in the bosonic case and anti-commute in fermionic case. We have the following properties to a given temperature $T=\beta^{-1}$,
\begin{eqnarray}
\bar{a}_{\beta}|0(\beta)\rangle = \bar{b}_{\beta}|0(\beta)\rangle=0.
\label{671878}
\end{eqnarray}
It follows that $\bar{a}_{\beta}$ and $\bar{b}_{\beta}$, given by equations (\ref{lyiul2}) and (\ref{lyiul4}), are annihilation operators to the thermofield vacuum and justify the term vacuum. On the other hand, $\bar{a}_{\beta}^{\dagger }$ and $\bar{b}_{\beta}^{\dagger }$, given by equations (\ref{lyiul1}) and (\ref{lyiul3}), excite the thermofield vacuum generating excited thermofield states. Since the Bogoliubov transformation is canonical, the corresponding commutations for bosons or anticommutations for fermions are preserved \cite{chu}.

Although this discussion of TFD is restricted to the more simple formulation in terms of creation 
and annihilation operators \cite{umezawab}, an elaborate discussion could include more
specific treatments as based on Schwinger
operators\cite{ademir}, gauge fields\cite{ojima}, supersimmetry\cite{nakahara} or the approach on phase space\cite{santanab}. These 
can be useful in more specific situations, as, for instance, spin 1/2 particles in a lattice\cite{ademir}, Higgs mechanism \cite{ojima}
 or in the approach to Wigner functions\cite{santanab}. We also
 point out that, in more specific cases, $\bar{G}$ in the Bogoliubov transformation $e^{-i\bar{G}}$ and the thermofield
vacuum $|0(\beta)\rangle$ can be more elaborate than (\ref{adk}) and (\ref{bog}).

\begin{figure}[h]
\centering
\includegraphics[scale=0.7]{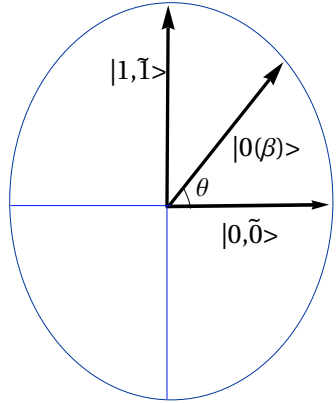}
\caption{ (Color online) Fermionic thermofield vacuum $|0(\beta)\rangle$, eq. (\ref{th}), in the plane generated by $|0,\tilde{0}\rangle$ and $|1,\tilde{1}\rangle$.}
\label{thermo}
\end{figure}

Now, as a simple example in this discussion, consider a fermionic oscillator, described by the hamiltonian $\hat{H}$
\begin{eqnarray}
\hat{H} &=& \omega \hat{a}^{\dagger}\hat{a}, \label{uslihl12341}\\
\tilde{H} &=& \omega \tilde{b}^{\dagger }\tilde{b}, \label{uslihl12342}
\end{eqnarray}
and the corresponding anticommutation relations
\begin{eqnarray}
\{\hat{a},\hat{a}^{\dagger }\} &=&\{\tilde{b},\tilde{b}^{\dagger }\}=1, \\
\{\hat{a},\hat{a}\} &=&\{\tilde{b},\tilde{b}\}=0, \\
\{\tilde{b}^{\dagger },\hat{a}^{\dagger }\} &=&\{\tilde{b},\hat{a}\}=0.
\end{eqnarray}
We can also define
\begin{eqnarray}
\bar{H} = \hat{H} - \tilde{H} = \omega \left( \hat{a}^{\dagger}\hat{a} - \tilde{b}^{\dagger }\tilde{b} \right).
\end{eqnarray}
This operator is hermitean $\bar{H}=\bar{H}^{\dagger}$ and gives the time evolution of an operator $\hat{A}$ by means of a Liouville-von Neumann equation \cite{ademir}. It also annihilates the thermofield vacuum at a given finite temperature $T=\beta^{-1}$
\begin{eqnarray}
\bar{H}|0(\beta)\rangle = 0.
\end{eqnarray}
The space $\mathcal{H}\otimes\tilde{\mathcal{H}}$ is generated, in this case, from the zero temperature vacuum and its excitations,
\begin{eqnarray}
|0,\tilde{0}\rangle &=& 1|0,\tilde{0}\rangle,  \label{00}\\
|1,\tilde{0}\rangle &=& \hat{a}^{\dagger }|0,\tilde{0}\rangle, \label{10} \\
|0,\tilde{1}\rangle &=&\tilde{b}^{\dagger }|0,\tilde{0}\rangle, \label{01} \\
|1,\tilde{1}\rangle &=& \hat{a}^{\dagger }\tilde{b}^{\dagger }|0,\tilde{0}\rangle. \label{11}
\end{eqnarray}
By applying the Bogolioubov transformation on the vacuum $|0,\tilde{0}\rangle$ we arrive at the following thermofield
vacuum
\begin{eqnarray}
|0(\beta )\rangle = \frac{1}{\sqrt{Z}}(|0,\tilde{0}\rangle +e^{\frac{%
-\beta \omega }{2}}|1,\tilde{1}\rangle).
\end{eqnarray}
From the normalization condition $|0(\beta )\rangle $, we derive the partition
function $Z=1+e^{-\beta \omega }$.
We then define
\begin{eqnarray}
u(\beta )&=&\frac{1}{\sqrt{1+e^{-\beta \omega }}},\\
v(\beta )&=&\frac{1}{\sqrt{1+e^{\beta \omega }}}.
\end{eqnarray}
Since the following relation is satisfied
\begin{eqnarray}
u(\beta )^{2}+v(\beta )^{2}=1,
\end{eqnarray}
we can also write
\begin{eqnarray}
u(\beta )&=&\cos \theta, \label{cs} \\
v(\beta )&=&\sin\theta, \label{ss}
\end{eqnarray}
where
\begin{eqnarray}
\theta =\tan ^{-1}(e^{-\frac{\beta \omega }{2}}).
\label{angulo}
\end{eqnarray}
In this case, the thermofield vacuum can be written as
\begin{equation}
|0(\beta )\rangle =\cos \theta |0,\tilde{0}\rangle +\sin \theta |1,\tilde{1}
\rangle .
\label{th}
\end{equation}
We can use this relation to calculate, for example, the mean value of the number operator
\begin{eqnarray}
\langle \hat{a}^{\dagger }\hat{a}\rangle =\langle 0(\beta )|\hat{a}^{\dagger }\hat{a}|0(\beta
)\rangle =\frac{e^{-\beta \omega }}{1+e^{-\beta \omega }},
\end{eqnarray}
which is the Fermi-Dirac distribution, where we have agreement with the statistical result as given in the equation (\ref{med}). We can also write \cite{umezawab}
\begin{eqnarray}
\langle \hat{a}^{\dagger }\hat{a}\rangle = \sin^{2}\theta.
\end{eqnarray}
The equation (\ref{th}) asserts that the fermionic thermofield vacuum is in the plane generated by $|0,\tilde{0}\rangle $ and $|1,\tilde{1}\rangle $. In fact, the action of the
Bogolioubov transformation on the vacuum excitations (\ref{00}), (\ref{10}), (\ref{01}) and (\ref{11}) is given by 
\begin{equation}
e^{-i\bar{G}}\left( 
\begin{array}{c}
|1,\tilde{1}\rangle  \\ 
|0,\tilde{0}\rangle 
\end{array}%
\right) =\left( 
\begin{array}{cc}
\cos \theta  & -\sin\theta  \\ 
\sin\theta  & \cos \theta 
\end{array}%
\right) \left( 
\begin{array}{c}
|1,\tilde{1}\rangle  \\ 
|0,\tilde{0}\rangle 
\end{array}%
\right)   \label{fermions1}
\end{equation}%
and 
\begin{equation}
e^{-i\bar{G}}\left( 
\begin{array}{c}
|1,\tilde{0}\rangle  \\ 
|0,\tilde{1}\rangle 
\end{array}%
\right) =\left( 
\begin{array}{cc}
1 & 0 \\ 
0 & 1%
\end{array}%
\right) \left( 
\begin{array}{c}
|1,\tilde{0}\rangle  \\ 
|0,\tilde{1}\rangle 
\end{array}%
\right) .  \label{fermions2}
\end{equation}%
It follows that the fermionic thermofield vacuum, eq. (\ref{th}), is in the plane generated by $|0,\tilde{0}%
\rangle $ and $|1,\tilde{1}\rangle $ and it corresponds to a rotation of $
\theta $, relatively to $|0,\tilde{0}\rangle $. On the other hand, the
action of the Bogolioubov transfomation on (\ref{10}) and (\ref{01}) has no effect, being equivalent to an identity operator in the plane generated by $|1,\tilde{0}%
\rangle $ and $|0,\tilde{1}\rangle $ (see figure \ref{thermo}).

Applying the thermofield operators (\ref{lyiul2}) and (\ref{lyiul4}) on (\ref{th}), considering the inverse of the Bogolioubov transformation
\begin{equation}
e^{i\bar{G}}\left( 
\begin{array}{c}
|1,\tilde{1}\rangle  \\ 
|0,\tilde{0}\rangle 
\end{array}%
\right) =\left( 
\begin{array}{cc}
\cos \theta  & \sin\theta  \\ 
-\sin\theta  & \cos \theta 
\end{array}%
\right) \left( 
\begin{array}{c}
|1,\tilde{1}\rangle  \\ 
|0,\tilde{0}\rangle 
\end{array}%
\right),   \label{fermions12}
\end{equation}%
we can show explicitly the thermofield vaccuum annihilations (\ref{671878})
\begin{eqnarray}
\bar{a}_{\beta}|0(\beta )\rangle &=& e^{-i\bar{G}}\hat{a}e^{i\bar{G}}\left(\cos \theta |0,\tilde{0}\rangle +\sin \theta |1,\tilde{1}
\rangle\right)\nonumber \\
 &=& e^{-i\bar{G}}\hat{a}\left(\cos^{2} \theta |0,\tilde{0}\rangle -\cos\theta\sin \theta |1,\tilde{1}\rangle + \cos\theta\sin \theta |1,\tilde{1}\rangle  + \sin^{2} \theta |0,\tilde{0}
\rangle\right) \nonumber \\
&=& 0,
\end{eqnarray}
and
\begin{eqnarray}
\bar{b}_{\beta}|0(\beta )\rangle &=& e^{-i\bar{G}}\tilde{b}e^{i\bar{G}}\left(\cos \theta |0,\tilde{0}\rangle +\sin \theta |1,\tilde{1}
\rangle\right)\nonumber \\
 &=& e^{-i\bar{G}}\tilde{b}\left(\cos^{2} \theta |0,\tilde{0}\rangle -\cos\theta\sin \theta |1,\tilde{1}\rangle + \cos\theta\sin \theta |1,\tilde{1}\rangle  + \sin^{2} \theta |0,\tilde{0}
\rangle\right) \nonumber \\
&=& 0.
\end{eqnarray}
Similar calculations could be done to the case of a bosonic oscillator, case where the thermofield vacuum is expressed by
\begin{eqnarray}
|0(\beta)\rangle = \sqrt{1-e^{-\beta\omega_{0}}}\sum_{n}e^{-\frac{n}{2}\beta \omega_{0}}|n,\tilde{n}\rangle,
\end{eqnarray}
with mean value of the number operator leading to a Bose-Einstein distribution.

\section{Thermofield vacuum and von Neumann entropy}

It is also interesting to consider the relation among the thermofield vacuum $ |0(\beta)\rangle \in \mathcal{H}\otimes\tilde{\mathcal{H}}$, its associated 
density operator $\hat{\rho}$ acting on the space $\mathcal{H}$ and the von Neumann entropy $S=S(\hat{\rho})$. 

\begin{figure}[h]
\centering
\includegraphics[scale=1.0]{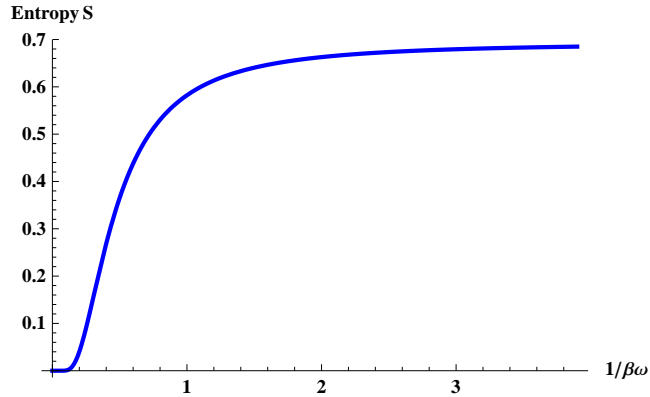}
\caption{(Color online) von Neumann entropy $S$ as a function of temperature scaled $T/\omega = 1/\beta\omega$.}
\label{gfr}
\end{figure}

As we discussed in the last section, by means of a Bogoliubov transformation in the state $|0,\tilde{0}\rangle$, a thermofield vacuum $|0(\beta)\rangle$ 
is generated.
This state is maximally entangled state in the space $\mathcal{H}\otimes\tilde{\mathcal{H}}$ \cite{ademir}. 
As a consequence it is non-factorable and, for this reason, at finite temperature, the thermofield vacuum is always 
associated to a mixed state in $\mathcal{H}$. 
The fact of $|0(\beta)\rangle$ be related to a mixed state in $\mathcal{H}$ is an important feature to be considered. 
Indeed, superpositions in the space $\mathcal{H}$ have a different role 
from superpositions states in the space $\mathcal{H}\otimes\tilde{\mathcal{H}}$, because 
in this higher space the effect of non-separability will reflect
in $\mathcal{H}$ the existence of $\tilde{\mathcal{H}}$. As an illustration, consider the following superposition state
\begin{eqnarray}
|\psi\rangle = \cos\theta |0\rangle + \sin\theta |1\rangle,
\label{sis}
\end{eqnarray}
where $\theta$ is given by (\ref{angulo}).
The density matrix associated to (\ref{sis}) is given by
\begin{eqnarray}
\hat{\rho}_{|\psi\rangle} = \left( 
\begin{array}{cc}
\cos^{2}\theta & \frac{1}{2} \sin 2\theta \\ 
\frac{1}{2} \sin 2\theta & \sin^{2}\theta
\end{array}%
\right) 
\label{sis2}
\end{eqnarray}
We can calculate the von Neumann entropy for the state (\ref{sis}) by means of its associate density matrix (\ref{sis2})
\begin{eqnarray}
S(\hat{\rho}_{|\psi\rangle})=-Tr(\hat{\rho}_{|\psi\rangle}\ln(\hat{\rho}_{|\psi\rangle}))=0.
\end{eqnarray}
This means that it is a pure state \cite{wei}$^{,}$\cite{tanabe}$^{,}$\cite{vedral}. This could also be verified from a more simple relation
\begin{eqnarray}
\hat{\rho}_{|\psi\rangle}(1-\hat{\rho}_{|\psi\rangle})=0.
\end{eqnarray}
On the other hand, the density matrix corresponding to the thermofield vacuum of the fermionic oscillator, eq. (\ref{th}),
\begin{equation}
|0(\beta )\rangle =\cos \theta |0,\tilde{0}\rangle +\sin \theta |1,\tilde{1}
\rangle 
\label{th6829}
\end{equation}
is associated to the density matrix
\begin{eqnarray}
\hat{\rho}=\frac{1}{Z}e^{-\beta \omega \hat{a}^{\dagger}\hat{a}}.
\end{eqnarray}
Then, if we calculate the von Neumann entropy for this state, we have
\begin{eqnarray}
S(\hat{\rho})&=&-Tr(\hat{\rho}\ln(\hat{\rho})) \nonumber \\
&=& \beta\omega \frac{e^{-\beta \omega }}{1+e^{-\beta \omega }} + \ln(1+e^{-\beta \omega}).
\label{stv}
\end{eqnarray}
This is a non-zero entropy and then the thermofield vacuum, a state in $\mathcal{H}\otimes\tilde{\mathcal{H}}$, 
is in fact associated to a mixed state in $\mathcal{H}$. In a more general case, the thermofield vacuum is associated to the following von Neumann 
entropy
\begin{eqnarray}
S(\hat{\rho}) &=& \beta\texttt{ } Tr(\hat{\rho} \hat{H}) + \ln(Z),
\label{entvact}
\end{eqnarray}
corresponding, at finite temperature, to a mixed state. When the temperature is zero, we have the pure case, 
where the thermofield vacuum is at zero temperature $|0,\tilde{0}\rangle$ and the entropy vanishes. In the extreme case 
where the temperature goes to infinity, the thermofield vacuum becames factorable $|1,\tilde{1}\rangle$, but in this case the entropy 
is $ S(\hat{\rho})= \ln(2)\approx 0.69$, the maximum value achieved by the entropy (see figure \ref{gfr}).

Then, in the space $\mathcal{H}\otimes\tilde{\mathcal{H}}$ generated in TFD, superpositions are associated to density matrices in 
the Hilbert space $\mathcal{H}$. A non-factorable state in $\mathcal{H}\otimes\tilde{\mathcal{H}}$ is associated to a mixed state in
$\mathcal{H}$. From quantum optics, it is known that two-mode squeezing operators lead to generation of 
an entangled state when acting in a two-mode vacuum\cite{vedral}. This is the case of the Bogoliubov 
transformation acting in the zero temperature vacuum $|0,\tilde{0}\rangle \in \mathcal{H}\otimes\tilde{\mathcal{H}}$ that leads to the thermofield
vacuum $|0(\beta)\rangle \in \mathcal{H}\otimes\tilde{\mathcal{H}}$.

\section{No-cloning theorem}

Suppose a cloning machine in a state $|A_{0}\rangle$ and a state that we wish to clone, e.g., the state $|n\rangle$. Then, the cloning process can be represented by
\begin{eqnarray}
|A_{0}\rangle|n\rangle \rightarrow |A_{n}\rangle|n,n\rangle
\label{c1}
\end{eqnarray}
where $|A_{n}\rangle$ is the final state of the machine and $|n,n\rangle$ is the final state where now we have two states $|n\rangle$.

In this process the Hilbert space generated by the Fock state $|n\rangle$ is increased to a product space generated by $|n,n\rangle$.

According to quantum mechanics this quantum cloning operation must be linear and unitary. Thus, if the state to clone is the following superposition
\begin{eqnarray}
z|n\rangle + w|m\rangle
\label{sup}
\end{eqnarray}
where $z,w \neq 0$, then
\begin{eqnarray}
|A_{0}\rangle\left(z|n\rangle + w|m\rangle \right) \rightarrow z|A_{n}\rangle|n,n\rangle + w|A_{m}\rangle|m,m\rangle,
\end{eqnarray}
where $|A_{n}\rangle$ and $|A_{m}\rangle$ are in general different states.

If $|A_{n}\rangle\neq|A_{m}\rangle$, then emergent system state of the cloning state is in a entangled state given by
\begin{eqnarray}
z|A_{n}\rangle|n,n\rangle + w|A_{m}\rangle|m,m\rangle
\end{eqnarray}
On the order hand, if $|A_{n}\rangle=|A_{m}\rangle$, then the emergent state is given by
\begin{eqnarray}
|A_{n}\rangle\left(z|n,n\rangle + w|m,m\rangle\right).
\label{c2}
\end{eqnarray}
Thus, in any case, the emergent state of the cloning machine never is a cloned state product given by
\begin{eqnarray}
\left(z|n\rangle + w|m\rangle\right)\left(z|n\rangle + w|m\rangle\right),
\label{sup2}
\end{eqnarray}
what proves the impossibility of a cloning machine. This result was proved in 1982 by Wooters and Zurek  \cite{wooters} and by Dieks in another way \cite{dieks}.

\section{No-cloning theorem in TFD}

\begin{figure}[h]
\centering
\includegraphics[scale=0.5]{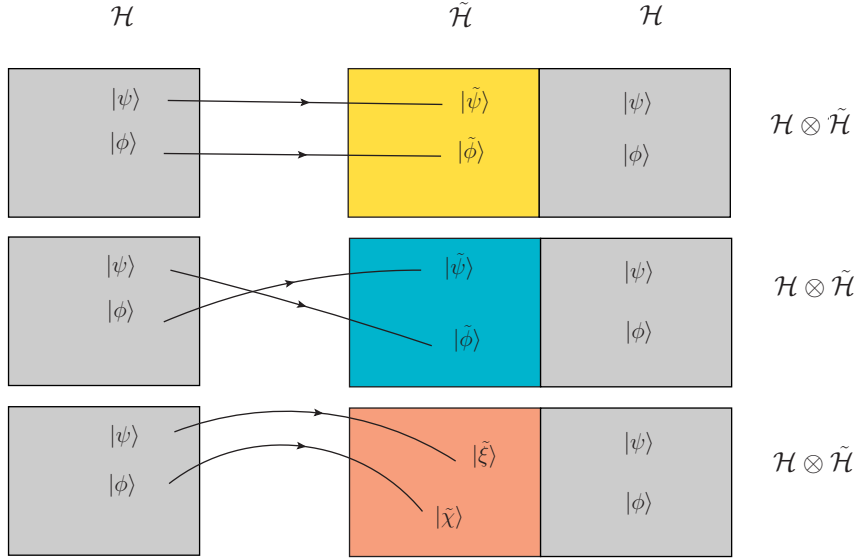}
\caption{(Color online) Possible doublings of freedom degrees of the Hilbert space $\mathcal{H}$ by the mapping $\mathcal{H} \rightarrow \mathcal{H}\otimes\tilde{\mathcal{H}}$.}
\label{doughbli}
\end{figure}

Now, we turn to the TFD procedure of doubling the degrees of freedom in the Hilbert space $\mathcal{H}$. This procedure can be summarized 
by the following mapping
\begin{eqnarray}
D_{TFD}: \mathcal{H} \rightarrow \mathcal{H}\otimes\tilde{\mathcal{H}}.
\label{dtfd}
\end{eqnarray}
This mapping works in this way: given a state $|\psi\rangle \in \mathcal{H}$, a state $|\tilde{\phi}\rangle \in \tilde{\mathcal{H}}$ is created 
such that $D_{TFD}$ can be expressed by
\begin{eqnarray}
D_{TFD}(|\psi\rangle)= |\psi\rangle\otimes|\tilde{\phi}\rangle.
\label{cdtfd}
\end{eqnarray}
There is no unique way of doubling the degrees of freedom of a Hilbert space (see figure \ref{doughbli}). As such, we need to make a choice. Indeed, given states $|\psi\rangle \in \mathcal{H}$, 
the doubling can be realized by means of a cloning procedure, where a cloning of $|\psi\rangle$ is created 
in the tilde space $\tilde{\mathcal{H}}$ and called $|\tilde{\psi}\rangle$. This mapping is expressed by
\begin{eqnarray}
|\psi\rangle \rightarrow |\psi\rangle|\tilde{\psi}\rangle,
\end{eqnarray}
In this case, $D_{TFD}$ is a cloning mapping and we can also write
\begin{eqnarray}
D_{TFD}(|\psi\rangle)= |\psi\rangle|\tilde{\psi}\rangle.
\end{eqnarray}
On the other hand, we can consider states $|\phi\rangle \in \mathcal{H}$ such that the doubling procedure works as a permutation, where 
states $|\psi\rangle  \in \mathcal{H}$ are lead to states $|\tilde{\phi}\rangle \in \tilde{\mathcal{H}}$  and 
states $|\phi\rangle  \in \mathcal{H}$ are lead to states $|\tilde{\psi}\rangle \in \tilde{\mathcal{H}}$. This mapping can be expressed by
\begin{eqnarray}
|\psi\rangle &\rightarrow& |\psi\rangle|\tilde{\phi}\rangle,\\
|\phi\rangle &\rightarrow& |\phi\rangle|\tilde{\psi}\rangle.
\end{eqnarray}
In this case, the mapping works as a permutation. In a more general case, the states $|\psi\rangle$ and $|\phi\rangle$ could match totally
 different states in the tilde space $\tilde{\mathcal{H}}$, as $|\tilde{\xi}\rangle$ and $|\tilde{\chi}\rangle$, leading to
\begin{eqnarray}
|\psi\rangle &\rightarrow& |\psi\rangle|\tilde{\xi}\rangle,\\
|\phi\rangle &\rightarrow& |\phi\rangle|\tilde{\chi}\rangle.
\end{eqnarray}
In terms of Fock states, the mapping (\ref{dtfd}) for a given state $|n\rangle$ can be written as $|n\rangle \rightarrow |n,\tilde{m}\rangle$.
Let us consider the strict case of a cloning mapping (\ref{cdtfd}), where Fock states are considered, such that from the state $|n\rangle$, we have
\begin{eqnarray}
|n\rangle \rightarrow |n,\tilde{n}\rangle.
\end{eqnarray}
For this case, we have
\begin{eqnarray}
D_{TFD}(|n\rangle)=|n,\tilde{n}\rangle.
\label{doubpure1}
\end{eqnarray}
Now, by considering a superposition state $z|n\rangle + w|m\rangle \in \mathcal{H}$, this doubling mapping can be written as
\begin{eqnarray}
z|n\rangle + w|m\rangle \in \mathcal{H} \rightarrow \left(z|n\rangle + w|m\rangle\right)\left(z|\tilde{n}\rangle + w|\tilde{m}\rangle\right) \in \mathcal{H}\otimes\tilde{\mathcal{H}}.
\label{dub}
\end{eqnarray}
or
\begin{eqnarray}
D_{TFD}(z|n\rangle + w|m\rangle)=\left(z|n\rangle + w|m\rangle\right)\left(z|\tilde{n}\rangle + w|\tilde{m}\rangle\right).
\label{doubpure2}
\end{eqnarray} 
The state $z|\tilde{n}\rangle + w|\tilde{m}\rangle$ corresponds to a copy of $z|n\rangle + w|m\rangle$ in the space $\tilde{\mathcal{H}}$. The complex
numbers $z$ and $w$ could go as complex conjugated in the tilde space, depending on our choice of the mapping. Let us first consider $z$ and $w$ real 
numbers.

From the no-cloning theorem exposed in the last section, we figure out that the cloning procedure in (\ref{cdtfd}) cannot be linear. 
In fact, if linearity
is validy here, we have
\begin{eqnarray}
D_{TFD}(z|n\rangle + w|m\rangle)= zD_{TFD}(|n\rangle) + wD_{TFD}(|m\rangle).
\end{eqnarray} 
It follows from (\ref{doubpure1}) that
\begin{eqnarray}
D_{TFD}(z|n\rangle + w|m\rangle)= z|n,\tilde{n}\rangle + w|m,\tilde{m}\rangle.
\label{doubpure3}
\end{eqnarray} 
However, by comparing (\ref{doubpure1}) and (\ref{doubpure3}) we arrive at
\begin{eqnarray}
z|n,\tilde{n}\rangle + w|m,\tilde{m}\rangle = z^{2}|n,\tilde{n}\rangle + w^{2}|m,\tilde{m}\rangle + wz\left(|m,\tilde{n}\rangle + |n,\tilde{m}\rangle\right).
\end{eqnarray}
This is only true in the case where $|z|=1$ and $|w|=0$  or $|z|=0$ and $|w|=1$ or both zero. If we had considered 
$z^{*}|\tilde{n}\rangle + w^{*}|\tilde{m}\rangle$ instead of $z|\tilde{n}\rangle + w|\tilde{m}\rangle$, where $z^{*}$ and $w^{*}$ are complex conjugated of
$z$ and $w$, the last equation would be
\begin{eqnarray}
z|n,\tilde{n}\rangle + w|m,\tilde{m}\rangle = |z|^{2}|n,\tilde{n}\rangle + |w|^{2}|m,\tilde{m}\rangle + \left(w^{*}z|m,\tilde{n}\rangle + z^{*}w|n,\tilde{m}\rangle\right).
\label{ndulhsd}
\end{eqnarray}
Again, this is only true in the case $|z|=1$ and $|w|=0$  or $|z|=0$ and $|w|=1$ or both zero. 

As a consequence $D_{TFD}$ in the cloning procedure form (\ref{cdtfd}) cannot be a linear mapping. 
This leads to a more strong result: We cannot devise an experiment by means of linear operations that lead to a doubling procedure in TFD 
to an arbitrary state $z|n\rangle + w|m\rangle$.

As the states $\left(z|n\rangle + w|m\rangle\right)\left(z|\tilde{n}\rangle + w|\tilde{m}\rangle\right)$ are factorable in $\mathcal{H}\otimes\tilde{\mathcal{H}}$ and
cannot result of a linear operation, we have as a particular consequence, they do not result from unitary evolution or any linear evolution. 
This can give us some route to deal with non-factorable states in $\mathcal{H}\otimes\tilde{\mathcal{H}}$ corresponding to thermofield states 
associated to physical states in $\mathcal{H}$. This result is of practical interest because a non-factorable state in $\mathcal{H}\otimes\tilde{\mathcal{H}}$ 
acts as an entanglement between $\mathcal{H}$ and $\tilde{\mathcal{H}}$, consequently cannot be associated to 
a pure state in $\mathcal{H}$, but only to a mixed state in $\mathcal{H}$.

However, the above result is not too much serious in the derivation of the thermofield vacuum. Indeed, a recipe to construct 
the thermofield vacuum is given by the following steps
\begin{eqnarray}
|0\rangle & \rightarrow & |0,\tilde{0}\rangle \label{cltfd} \\
|0,\tilde{0}\rangle & \rightarrow & |0(\beta)\rangle = e^{-i\bar{G}}|0,\tilde{0}\rangle. \label{bogtfd}
\end{eqnarray}
The doubling procedure in (\ref{cltfd}) corresponds to the case $|z|=1$, $|w|=0$ and $n=0$, where the equality (\ref{ndulhsd}) is true.
In the step (\ref{cltfd}) we have a separable state in the space $\mathcal{H}\otimes\tilde{\mathcal{H}}$, i.e.,
 we can distinguish $|0\rangle \in \mathcal{H}$ and $|\tilde{0}\rangle \in \tilde{\mathcal{H}}$. 
In the step (\ref{bogtfd}), the state is not separable at finite temperatures and we cannot divide $|0(\beta)\rangle \in \mathcal{H}\otimes\tilde{\mathcal{H}}$ 
in a part pertaining to the space $\mathcal{H}$ and other pertaining to $\tilde{\mathcal{H}}$. In this sense, the thermofield vacuum is entangled 
in the space $\mathcal{H}\otimes\tilde{\mathcal{H}}$.

It is important to emphasize the differences between the spaces $\mathcal{H}\otimes\tilde{\mathcal{H}}$ and $\mathcal{H}\otimes\mathcal{H}$. 
The space $\mathcal{H}\otimes\mathcal{H}$ corresponds to product space of two Hilbert spaces corresponding both to quantum physical systems. 
On the other hand,  $\mathcal{H}\otimes\tilde{\mathcal{H}}$ comes from the TFD  procedure of doubling the freedom degrees in a Hilbert space 
$\mathcal{H}$. 

We can also discuss the no-cloning theorem in context of thermofield states in TFD. Excitations from the thermofield vacuum are also
existing states in the Hilbert space $\mathcal{H}\otimes\tilde{\mathcal{H}}$. Let us consider the states
\begin{eqnarray}
|1(\beta)\rangle = \bar{a}_{\beta}^{\dagger}|0(\beta )\rangle
\label{1beta1}
\end{eqnarray}
and
\begin{eqnarray}
|\tilde{1}(\beta)\rangle = \bar{b}_{\beta}^{\dagger}|0(\beta )\rangle,
\label{1beta2}
\end{eqnarray}
corresponding to the action of thermofield creation operators $\bar{a}_{\beta}^{\dagger}$, equation (\ref{lyiul1}), and $\bar{b}_{\beta}^{\dagger}$, 
equation (\ref{lyiul3}), in the thermofield vacuum. If we consider again the case of a fermionic oscillator
\begin{equation}
|0(\beta )\rangle =\cos \theta |0,\tilde{0}\rangle +\sin \theta |1,\tilde{1}\rangle,
\end{equation}
Then, the thermofield states (\ref{1beta1}) and (\ref{1beta2}) will be explicitly written as
\begin{eqnarray}
|1(\beta)\rangle &=& e^{-i\bar{G}}\hat{a}^{\dagger }e^{i\bar{G}}\left(\cos \theta |0,\tilde{0}\rangle +\sin \theta |1,\tilde{1}\rangle\right) \nonumber\\
&=& e^{-i\bar{G}}\hat{a}^{\dagger }\left(\cos^{2} \theta |0,\tilde{0}\rangle -\cos\theta\sin \theta |1,\tilde{1}\rangle + \cos\theta\sin \theta |1,\tilde{1}\rangle  + \sin^{2} \theta |0,\tilde{0}
\rangle\right)\nonumber\\
&=& e^{-i\bar{G}}\hat{a}^{\dagger }|0,\tilde{0}\rangle\nonumber\\
&=& e^{-i\bar{G}}|1,\tilde{0}\rangle\nonumber\\
&=& |1,\tilde{0}\rangle,
\end{eqnarray}
and
\begin{eqnarray}
|\tilde{1}(\beta)\rangle &=& e^{-i\bar{G}}\tilde{b}^{\dagger }e^{i\bar{G}}\left(\cos \theta |0,\tilde{0}\rangle +\sin \theta |1,\tilde{1}\rangle\right)\nonumber\\
&=& e^{-i\bar{G}}\tilde{b}^{\dagger }\left(\cos^{2} \theta |0,\tilde{0}\rangle -\cos\theta\sin \theta |1,\tilde{1}\rangle + \cos\theta\sin \theta |1,\tilde{1}\rangle  + \sin^{2} \theta |0,\tilde{0}
\rangle\right)\nonumber\\
&=& e^{-i\bar{G}}\tilde{b}^{\dagger }|0,\tilde{0}\rangle\nonumber\\
&=& e^{-i\bar{G}}|0,\tilde{1}\rangle\nonumber\\
&=& |0,\tilde{1}\rangle. 
\end{eqnarray}
Note that, in this simple case, the thermofield excitations $|1(\beta)\rangle$ and $|\tilde{1}(\beta)\rangle$ are factorable 
in $\mathcal{H}\otimes\tilde{\mathcal{H}}$, although the thermofield vacuum does not. On the other hand,
\begin{eqnarray}
w|1(\beta)\rangle + z|\tilde{1}(\beta)\rangle &=& w|1,\tilde{0}\rangle + z|0,\tilde{1}\rangle,
\end{eqnarray}
is a possible thermofield superposition state in 
$\mathcal{H}\otimes\tilde{\mathcal{H}}$ and corresponds to a non-factorable state. Other examples of superpositions can also come. 
For example, a combination of the thermofield vacuum and the excitation (\ref{1beta1}),
\begin{eqnarray}
u|0(\beta)\rangle + v|1(\beta)\rangle
\label{sab}
\end{eqnarray}
where we can impose normalization
\begin{eqnarray}
|u|^{2} + |v|^{2}=1.
\end{eqnarray}
If we can clone the state (\ref{sab}), the cloned state will appear 
as an state in the product space $\left(\mathcal{H}\otimes\tilde{\mathcal{H}}\right)\otimes\left(\mathcal{H}\otimes\tilde{\mathcal{H}}\right)$ and 
the resulting state will be
\begin{eqnarray}
\left(u|0(\beta)\rangle + v|1(\beta)\rangle\right)\left(u|0(\beta)\rangle + v|1(\beta)\rangle\right) \in \left(\mathcal{H}\otimes\tilde{\mathcal{H}}\right)\otimes\left(\mathcal{H}\otimes\tilde{\mathcal{H}}\right).
\end{eqnarray}
In this sense, the cloning procedure for thermofield states is a mapping from a twofold to a fourfold space, i.e.,
\begin{eqnarray}
\left(\mathcal{H}\otimes\tilde{\mathcal{H}}\right) \rightarrow \left(\mathcal{H}\otimes\tilde{\mathcal{H}}\right)\otimes\left(\mathcal{H}\otimes\tilde{\mathcal{H}}\right)
\label{clonntfd}
\end{eqnarray}
Again, we can derive consequences of the no-cloning theorem. Let us call the cloning mapping in (\ref{clonntfd}) by $C_{TFD}$, 
\begin{eqnarray}
C_{TFD}:\left(\mathcal{H}\otimes\tilde{\mathcal{H}}\right) \rightarrow \left(\mathcal{H}\otimes\tilde{\mathcal{H}}\right)\otimes\left(\mathcal{H}\otimes\tilde{\mathcal{H}}\right)
\end{eqnarray}
such that
\begin{eqnarray}
C_{TFD}\left(u|0(\beta)\rangle + v|1(\beta)\rangle\right)&=& \left(u|0(\beta)\rangle + v|1(\beta)\rangle\right)\left(u|0(\beta)\rangle + v|1(\beta)\rangle\right), \label{uikas}\\
C_{TFD}\left(|0(\beta)\rangle\right)&=& |0(\beta)\rangle|0(\beta)\rangle,\\
C_{TFD}\left(|1(\beta)\rangle\right)&=& |1(\beta)\rangle|1(\beta)\rangle.
\end{eqnarray}
If this cloning is linear, it will happen
\begin{eqnarray}
C_{TFD}\left(u|0(\beta)\rangle + v|1(\beta)\rangle\right)= uC_{TFD}(|0(\beta)\rangle) + vC_{TFD}(|1(\beta)\rangle).
\end{eqnarray}
However, this will imply 
\begin{eqnarray}
C_{TFD}\left(u|0(\beta)\rangle + v|1(\beta)\rangle\right)= u|0(\beta)\rangle|0(\beta)\rangle + v|1(\beta)\rangle|1(\beta)\rangle.
\label{uikas2}
\end{eqnarray}
Compared to (\ref{uikas}), (\ref{uikas2}) is a wrong relation except to the cases $|u|=1$ and $|v|=0$ or $|u|=0$ and $|v|=1$ or both zero.
It results the cloning mapping $C_{TFD}$ is not linear. In fact, it cannot be antilinear (conjugate-linear) for the same reasons.

Although we have considered a specific superposition, as in the no-cloning theorem, the above result is also applied to a more general case.
Consequently, the cloning mapping $C_{TFD}$ cannot result of an unitary evolution applied to a general thermofield state. For a bosonic case, 
for example the bosonic oscillator, the thermofield vacuum has a 
more complicated form and in general thermofield excitations are non-factorable. In any case, the no-cloning result
keeps applying. 

As the no-cloning theorem has an important role in the case of quantum information protocols, this result for TFD can also be important for
quantum information protocols in this context.

\section{Conclusion}

Here we have discussed the no-cloning theorem in the approach of TFD. Given a Hilbert space $\mathcal{H}$ the no-cloning 
theorem imposes a restriction in the access to states in the Hilbert space $\mathcal{H}\otimes\mathcal{H}$, leading to a physical restriction 
in the procedure of copying states. In the case of TFD, we discussed the no-cloning theorem in 
the procedure of doubling a Hilbert space $\mathcal{H}$ by means of the construction of 
a Hilbert space $\mathcal{H}\otimes\tilde{\mathcal{H}}$ and its application to superpositions 
of thermofield states in TFD. By means of a doubling procedure using a cloning mapping it results that a doubling cannot be realized
by means of a linear mapping. This leads to the more strong result that we cannot devise an experiment by means of linear operations 
that lead to a doubling procedure in TFD 
to an arbitrary state $z|n\rangle + w|m\rangle$. 
In the case of thermofield states, 
the cloning procedure for thermofield states is a mapping from a twofold space $\left(\mathcal{H}\otimes\tilde{\mathcal{H}}\right)$ to a fourfold space 
$\left(\mathcal{H}\otimes\tilde{\mathcal{H}}\right)\otimes\left(\mathcal{H}\otimes\tilde{\mathcal{H}}\right)$. In this case, it also results that
the cloning mapping cannot be realized by means of linear or antilinear operations. In fact, an unitary evolution or any other linear mapping 
cannot lead to a cloning of an arbitrary superposition of thermofield states in TFD.

As no-cloning theorem has an important role in quantum information theory, the discussions here can motivate further studies 
of quantum information protocols\cite{buzek}$^{,}$\cite{gotesman}$^{,}$\cite{benet} using the formalism of TFD
\cite{das,umezawab,umezawab2,santanab}.

\section{Acknowledgments}

The author thanks CAPES (Brazilian government agency) for financial support, referees for some relevant references and suggestions, professors T. M. Rocha Filho and A. E. Santana for some previous discussions.

\vspace*{-5pt}   

\end{document}